\documentclass[a4paper,twocolumn,11pt,accepted=2021-11-17]{quantumarticle}
\pdfoutput=1
\usepackage[utf8]{inputenc}
\usepackage[english]{babel}
\usepackage[T1]{fontenc}
\usepackage{amsmath}
\usepackage{hyperref}

\usepackage{tikz}
\usepackage{lipsum}
\usepackage[numbers,sort&compress]{natbib}

\newcommand{\ket}[1]{\vert{#1}\rangle}

\newcommand{\outpr}[2]{\vert{#1}\rangle\langle{#2}\vert}

\newcommand{\proj}[1]{\outpr{#1}{#1}}

\begin{document}

\title{Efficient Characterization of Quantum Evolutions via a Recommender System}

\author{Priya Batra}
\orcid{}
\email{priya.batra@students.iiserpune.ac.in}
%\homepage{https://96ya.github.io/priyabtr/}
\thanks{}
\author{Anukriti Singh}
\email{anukriti.runjhun@gmail.com}
\author{T S Mahesh}
\affiliation{Department of Physics and NMR Research Center,\\
	Indian Institute of Science Education and Research, Pune 411008, India}
\email{mahesh.ts@iiserpune.ac.in}

\maketitle

\begin{abstract}
 We demonstrate characterizing quantum evolutions via matrix factorization algorithm, a particular type of the recommender system (RS).  A system undergoing a quantum evolution can be characterized in several ways.  Here we choose (i) quantum correlations quantified by measures such as entropy, negativity, or discord, and (ii) state-fidelity.
 Using quantum registers with up to 10 qubits, we demonstrate that an RS can efficiently characterize both unitary and nonunitary evolutions.  After carrying out a detailed performance-analysis of the RS in two-qubits, we show that it can be used to distinguish a clean database of quantum correlations from a noisy or a fake one.  Moreover, we find that the RS brings about a significant computational advantage for building a large database of quantum discord, for which no simple closed-form expression exists. Also, RS can efficiently characterize systems undergoing nonunitary evolutions in terms of quantum discord reduction as well as state-fidelity. Finally, we utilize RS for the construction of discord phase space in a nonlinear quantum system.
\end{abstract}

\section{Introduction}

Machine learning is increasingly featuring in almost every aspect of our understanding of the world. A popular class of machine learning, namely 
the Recommender System (RS), is often used to orient consumers towards certain products according to individual preferences \cite{1167344,funk2006netflix,Schafer2001}.
There are many approaches to build an RS, such as the \textit{content based system}, where recommendations are based on users' past experience \cite{10.1145/258549.258592,MAES1995811}, \textit{knowledge based system}, where one uses the knowledge of users and items to render the recommendations \cite{burke2000knowledge}, and the widely used \textit{collaborative filtering}, which exploits the user-item correlation, i.e., the interconnection between users' preferences among products with the recommendations provided by other users \cite{10.1145/138859.138867, Schafer2007}.
Collaborative filtering can be implemented by neighborhood methods or by latent-factor modeling. Matrix factorization algorithm (MFA) is a popular tool for implementing latent-factor modeling based RS. It can predict the rating of a specific item by a particular user based on the self-rating of other items as well as others' ratings of various items  \cite{5197422}.

The recent spurt of machine learning applications for quantum information tasks includes its usage in quantum tomography \cite{torlai2018neural, PhysRevA.96.062327}, quantum error correction \cite{PhysRevLett.119.030501}, quantum control \cite{PhysRevA.95.012335}, understanding quantum phase transitions \cite{carrasquilla2017machine,PhysRevB.100.045129}, and studying quantum-many-body problems \cite{doi:10.1126/science.aag2302, gao2017efficient}. It has been shown recently that machine learning techniques can work as state classifiers too. Sirui Lu and co-workers have shown the separability criteria of entangled state using convex hull approximation and supervised learning \cite{PhysRevA.83.012327}. Ma and Yung showed that it is possible to classify the separable and entangled states using artificial neural networks \cite{ma2018transforming}. The work has been further extended to experimental data \cite{PhysRevLett.120.240501} and later has been applied to simultaneous learning of multiple nonclassical correlations as well \cite{PhysRevLett.123.190401}. Valeria Cimini and co-workers proposed an artificial neural network to calculate the negativity of the Wigner function for multi-mode quantum states \cite{PhysRevLett.125.160504}. The connection between geometric and entropy discord has also been  explored using machine learning \cite{Li_2019}.

In this work, we employ an RS for characterizing quantum evolutions in terms of change in quantum correlations as well as fidelities.  Using a two-qubit register, we first perform a detailed study of the RS performance with respect to prediction accuracy, dependence on the database dimensions, and dependence on the size of latent vectors.     By observing the dependence of prediction efficiency on systematically introduced noise in the input database, we infer that an RS can identify a noisy (or fake) database from a genuine one. Using two- and three-qubit registers, we compare the computational efficiency of RS prediction of quantum discord with that of the standard method.  We demonstrate that, within certain precision limits, an RS can be considerably faster than the standard method.  For example, starting from a sparse  database of quantum discord, RS can completely fill it out an order of magnitude faster than the standard methods.  We then show the scalability of RS in larger systems by predicting quantum correlations and fidelities of unitary evolutions on registers with up to 10 qubits.  We also examine the RS ratings of nonunitary  evolutions by predicting discord changes and fidelities of a two-qubit system subjected to independent single-qubit decoherence channels. 
Finally, we demonstrate another important application of RS prediction in studying quantum nonlinear systems.  Specifically, we use the RS prediction to efficiently construct the phase space diagram of a quantum kicked top.

The paper is organized as follows. After briefly introducing the RS via MFA in Sec. \ref{MF}, we describe adapting it to characterize quantum evolutions via quantum-correlation changes in Sec. \ref{adapt}. 
We demonstrate the rating of unitary evolutions in Sec. \ref{unitarysec} and of nonunitary evolutions in Sec. \ref{nonunitarysec}. We then describe the RS prediction of quantum discord phase space in sec. \ref{sec:application}. 
Finally we conclude in Sec. \ref{summary}.

\section{Recommender system via matrix factorization}
\label{MF}
\begin{figure}[t]
	\centering
	\includegraphics[trim=1.3cm 2.6cm 0.6cm 0.3cm,width=8.5cm,clip=]{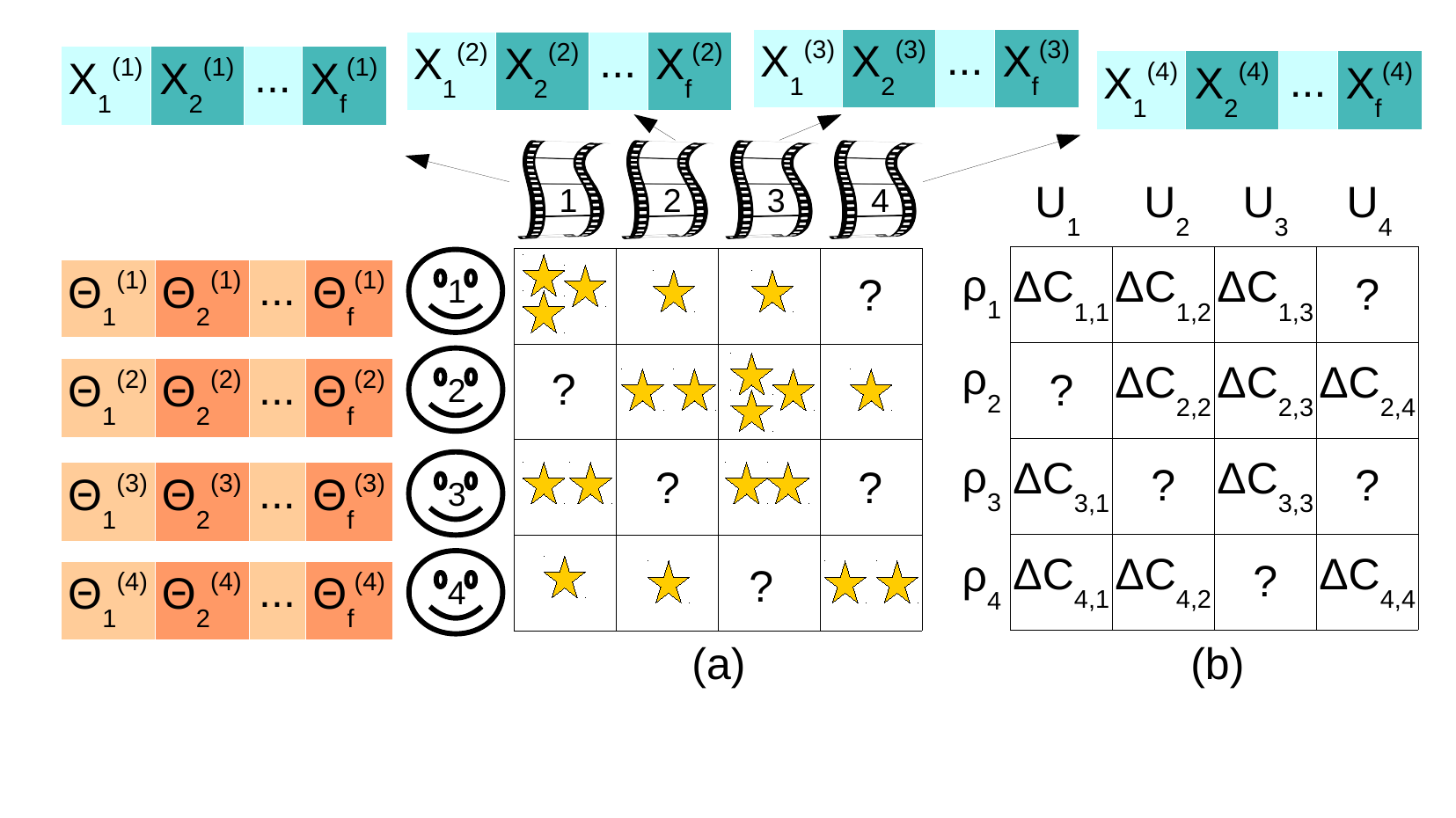}
	\caption{(a) Movie database. (b) Database of change $\Delta C_{i,j}$ in quantum correlations of states $\rho_i$ caused by  evolutions $U_j$. }
	\label{ratingscheme}
\end{figure}

MFA represents the user-item interaction in a lower dimensional latent space \cite{wiki:mfa}.
Consider a set of $m$ users and a set of $n$ items.	Each user $i$ is represented using a parameter vector $\Theta^{(i)}  \in \mathbbm{R}^f $ and each item $j$ is represented using a feature vector $X^{(j)} \in \mathbbm{R}^f$ (Fig. \ref{ratingscheme}(a)). Here $\mathbbm{R}^f$ is the coordinate space of dimension $f$ over real numbers.   
The interaction between a user $i$ and an item $j$ is modeled by the scalar product
\begin{equation}
	r_{i,j} = \Theta^{(i)} \cdot X^{(j)} = \sum_{l=1}^f \Theta^{(i)}_l  X^{(j)}_l
\end{equation}
that is conceived as the predicted rating. Now the task reduces to finding for all users and all items, the latent vectors consistent with known ratings and thereby making the best predictions about the unknown elements.

We start with random guesses for the latent vectors and evaluate rating elements $r_{i,j}$.  Let $\kappa = \{(i,j)\}$ be the set of user-item pairs for which the ratings ${\cal R}_{i,j}$ are known. The mismatch between the evaluated and actual ratings is quantified by
\begin{equation}
	J_0 = \sum_{(i,j)\in \kappa } (r_{i,j} - {\cal R}_{i,j})^2.
\end{equation}
The next step is to optimize the latent vectors by minimizing the mismatch $J_0$.
Over-fitting is avoided by including two regularization terms in the objective function
\begin{equation}
	J = \frac{J_0}{2} 
	+ \frac{\lambda}{2} \sum_{i=1}^{m} \Vert \Theta^{(i)} \Vert + \frac{\lambda}{2} \sum_{j=1}^{n} \Vert X^{(j)} \Vert, 
\end{equation} 
where $\lambda$ is the regularization parameter and $\Vert \cdot \Vert$ denotes the norm of the vector. 
We may now use the first-order gradient descent algorithm for the minimization task. The gradients in the $k$th iteration can be cast as 
\begin{align}
	G_{\Theta^{(i,k)}} &= \frac{\partial J^{(k)}}{\partial \Theta^{(i,k)}} 
	\nonumber \\
	&= \sum_{j \in \kappa_{(i,.)}} (r_{i,j}^{(k)} - {\cal R}_{i,j})  X^{(j,k)} +\lambda \Theta^{(i,k)}, \nonumber      \\
	G_{X^{(j,k)}} &= \frac{\partial J^{(k)}}{\partial X^{(j,k)}} \nonumber \\
	&=  \sum_{i \in \kappa_{(.,j)}} (r_{i,j}^{(k)} - {\cal R}_{i,j} )  \Theta^{(i,k)} +\lambda X^{(j,k)},
\end{align} 
where $\kappa_{(i,.)}, \kappa_{(.,j)} \subset \kappa$ with the respective indices being fixed.
The latent vectors for $(k+1)$th iteration are now updated according to
\begin{align}
	\Theta^{(i,k+1)} &= \Theta^{(i,k)} - \alpha G_{\Theta^{(i,k)}} ~\mbox{and} \nonumber \\
	X^{(j,k+1)} &= X^{(j,k)} - \alpha G_{X^{(j,k)}},
\end{align}
where $\alpha$ is the suitable step size.
After a final number $K$ of iterations, with a desired value of the objective function, one can calculate the final rating $r_{i,j}^{(K)} = \Theta^{(i,K)} \cdot X^{(j,K)}$ for all unknown elements $(i,j) \notin \kappa$ \cite{wiki:mfa,coficode}.

\section{Adapting the recommender system}
\label{adapt}
It is insightful to consider the example of  viewers' ratings of movies. Every viewer rates some of the movies, thereby leaving an imprint of personal tastes or preferences, apart from assessing individual movies (Fig. \ref{ratingscheme}(a)). 
The RS aims to predict ratings for movies not yet seen/rated by the viewer, considering the viewer's tastes as well as the recommendations provided by other viewers.  In our RS, quantum states are viewers and quantum evolutions are movies (Fig. \ref{ratingscheme}(b)). A state transformed by a quantum evolution undergoes a change in its internal quantum correlation that can be labeled as the state's rating of the evolution. Our objective is to predict unknown ratings in the state-evolution database.

Consider a database ${\cal R}$ formed by ${n_s}$ randomly generated quantum states $\{\rho_1,\rho_2,\cdots,\rho_{n_s}\}$  and ${n_u}$ randomly generated unitary operators $\{U_1,U_2,\cdots,U_{n_u}\}$.  A random unitary operator is generated by matrix exponentiation of a random anti-Hermitian generator. 
Random pure states are generated simply by normalizing a random vector of complex elements.
We use Bures method \cite{Al_Osipov_2010,maziero2015random} to generate a random mixed state
$\rho = R R^\dagger/\mbox{Tr}[RR^\dagger] ~\mbox{with}~  R = (\mathbbm{1}+U)A,$
where $\mathbbm{1}$ is the identity matrix, $U$ is a random unitary operator, and $A$ is a random complex matrix.
An element of the database matrix ${\cal R}$, corresponding to $i$th state and $j$th unitary operator is the change in a measure $C$ of quantum correlation,
\begin{equation}
	{\cal R}_{i,j} = \Delta C_{i,j} = C(U_j \rho_i U_j^\dagger) - C(\rho_i).
\end{equation}
We consider the following three measures of quantum correlation: 

(i) \textit{von Neumann Entropy}
$S(\rho^{A}_i) = -\mbox{Tr}[\rho^A_i\log\rho^A_i]$
of a pure state $\rho^{AB}_i = \proj{\psi^{AB}_i}$ \cite{Plenio2014}.
The entropy change $\Delta S_{i,j} = S(U_j \rho^A_i U^\dagger_j)-S(\rho^A_i)$ is listed in the database as entropy rating. 

(ii) \textit{Negativity} of a general quantum state $\rho^{AB}_i$, pure or mixed
$N(\rho^{AB}_i) = \frac{\Vert R_A \Vert-1}{2},$
where $R_A$ is the partial transpose of $\rho^{AB}$ with respect to the subsystem $A$, and $\Vert R_A \Vert = \mbox{Tr}\sqrt{R_A^\dagger R_A}$ is the trace norm of $R_A$.  The change in negativity brought about by a unitary operator $U_j$ is $\Delta N_{i,j} = N(U_j\rho^{AB}_iU_j^\dagger)-N(\rho^{AB}_i)$.  Since the two-qubit negativity is bounded between 0 and 1/2, it is convenient to compare twice of negativity with other measures.

(iii) \textit{Discord} 
$D(\rho^{AB}_i) = I(\rho^{AB}_i) - \mbox{max}_{\{\Pi^A\}} J(\rho^{AB}_i),$ where
$I(\rho^{AB}_i) = S(\rho^A_i) + S(\rho^B_i) - S(\rho^{AB}_i)$ and
$J(\rho^{AB}_i) = S(\rho^B_i) - S(\rho^B_i|\rho^A_i)$ are the classically equivalent measures of mutual information
\cite{PhysRevLett.88.017901, PhysRevA.86.012309}. Discord is estimated numerically by maximizing $J$ over all possible measurement bases $\{\Pi^A\}$ in subsystem $A$. Despite being a stronger measure of quantum correlation, it has no simple analytical expression unlike entropy and negativity. Yichen Huang has recently shown that the complexity for computing discord is NP complete and the computational resource for computing discord is set to grow exponentially with the dimension of the Hilbert space \cite{Huang_2014}. Therefore it is interesting to see how well machine learning performs in predicting the change in quantum discord $\Delta D_{i,j} = D(U_j\rho^{AB}_iU^\dagger_j) - D(\rho^{AB}_i)$. Of course, for an uncorrelated initial state, the prediction is an estimate of the discord content in the transformed state itself.

\section{Rating unitary evolutions}
\label{unitarysec}
\begin{figure}[t]
	\centering
	\includegraphics[trim=2cm 2cm 2cm 1.5cm,width=8cm,clip=]{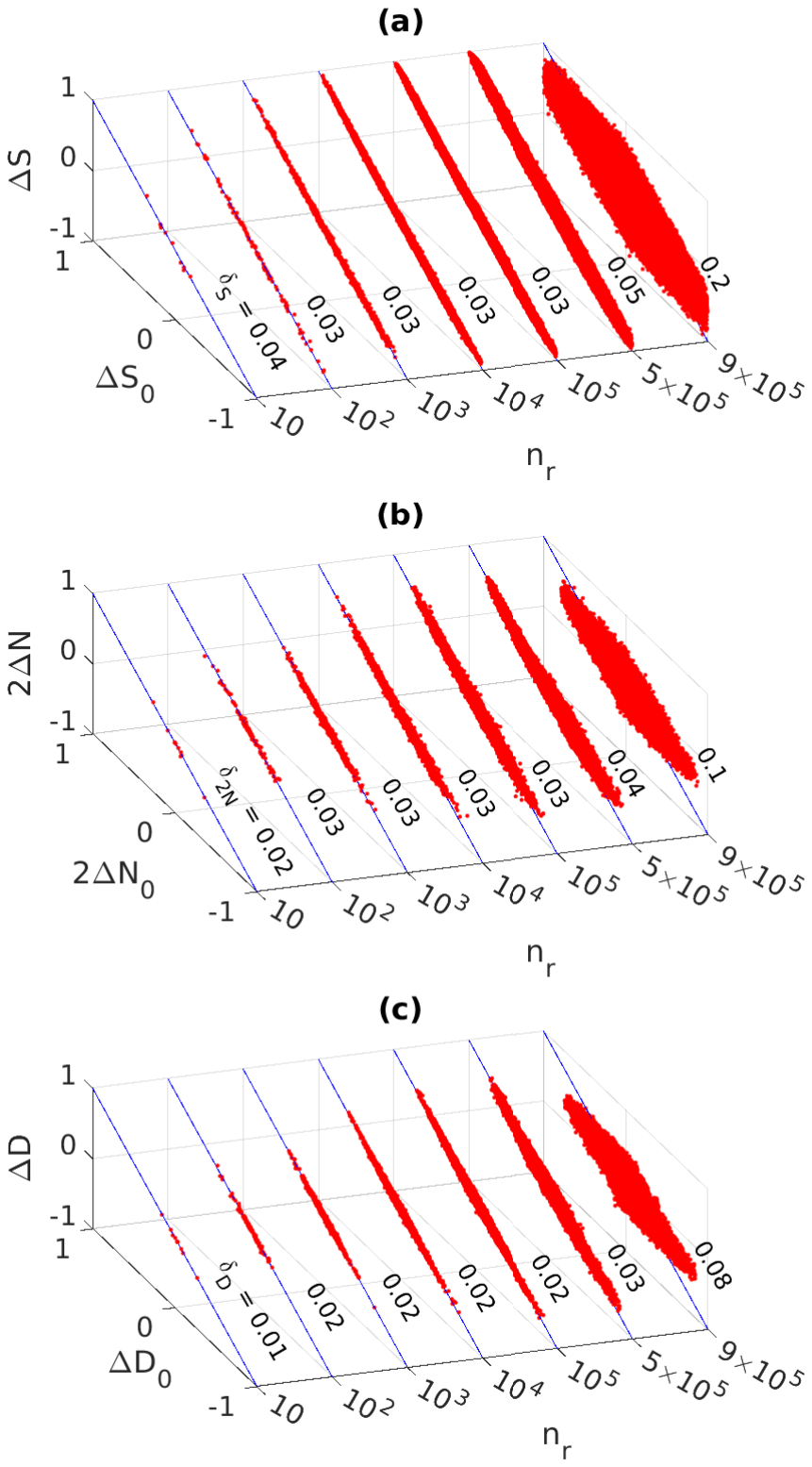}
	\caption{Red dots represent the RS predictions of changes in two-qubit correlations: entropy change $\Delta S$ (a), negativity change $2\Delta N$ (b), and discord change $\Delta D$ (c), plotted against the actual values ($\Delta S_0$, $2\Delta N_0$, $\Delta D_0$) and the number $n_r$ of unknown elements in a database of 1000 states and 1000 unitary operators.  Blue lines represent the ideal case $\Delta C_0=\Delta C$.  The RMSD value $\delta_C$ is also shown in each case.}
	\label{corr2qa}
\end{figure}

We first consider a million-element rating database ${\cal R}$ formed by ${n_s} = 1,000$ randomly generated two-qubit quantum states and ${n_u} = 1,000$ randomly generated two-qubit unitary evolutions. 
From the complete database, we randomly remove a set $\bar{\kappa} = \{(i,j)\}$ of $n_r$ elements, which are to be predicted by the RS.  The prediction error is measured by the root-mean-square deviation (RMSD)
\begin{equation}
	\delta_C = \left[\sum_{(i,j)\in \bar{\kappa}} (\Delta C_{0_{i,j}}-\Delta C_{i,j})^2 \right]^{1/2},
\end{equation}
between the actual correlation changes $\{\Delta C_{0_{i,j}}\}$ and their predicted values $\{\Delta C_{i,j}\}$.  
Results of the RS predictions for a two-qubit register are displayed in Fig. \ref{corr2qa} (a-c). Here, the predicted values $\Delta C$ are plotted versus the actual values $\Delta C_0$ for various numbers $n_r$ of unknown ratings.  Entropy ratings (Fig. \ref{corr2qa}(a)) are for pure states, while the negativity (Fig. \ref{corr2qa}(b)) and discord ratings (Fig. \ref{corr2qa}(c)) are for mixed states.  It is clear that the RS is quite successful in predicting the changes in all the correlation measures. 
Particularly, the discord predictions are impressive, and even better than that of other correlations.
The RMSD values remained below 0.05, except for $n_r = 9 \times 10^5$ which corresponds to 90\% of the million-element database being unknown.  
A systematic growth of RMSD w.r.t. $n_r$ is also observed, which is expected since, as the database becomes more and more sparse, the minimum in latent space turns shallower, and accordingly more uncertain will be the predictions.

\begin{figure}
	\centering
	\includegraphics[trim=4cm 11cm 4cm 11cm,width=7cm,clip=]{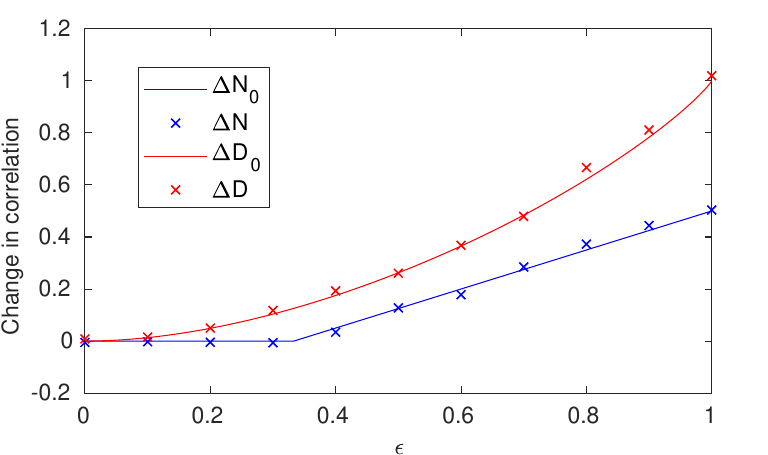}
	\caption{Symbols indicate the predicted values of $\Delta N$ and $\Delta D$ values for the quantum uncorrelated input state $\rho_-$ transformed into the Werner state $\rho_W$ by the CNOT operation. Solid lines represent actual values and symbols represent predicted values as indicated by the legend box.  The RMSD values are $\delta_{2N} = 0.026$ and $\delta_{D} = 0.038$.}
	\label{werner}
\end{figure}

\subsection{Rating the Werner state}
To observe the RS in action with a concrete example, we replace one of the random evolutions with the  two-qubit controlled-NOT operation
$U_{\mbox{CNOT}} = \proj{0} \otimes \mathbbm{1}_2 + \proj{1} \otimes \sigma_x,$
where $\mathbbm{1}_2$ is the $2\times 2$ identity operation and $\sigma_x$ is the NOT gate.  Additionally, we choose one of the input states to be the separable state,
$\rho_- = (1-\epsilon)\mathbbm{1}_4/4 + \epsilon \proj{-} \otimes \proj{1},$
where $\ket{-} = (\ket{0}-\ket{1})/\sqrt{2}$ is the single-qubit superposition state and the scalar quantity $\epsilon$ is the purity of state.  While the state $\rho_-$ is a pure state for $\epsilon = 1$ and becomes completely mixed for $\epsilon = 0 $, it is separable for all values of $\epsilon$, and hence has zero negativity and discord.  Upon acted by the CNOT operation, $\rho_-$ transforms into the Werner state
$\rho_W = (1-\epsilon)\mathbbm{1}_4/4 + \epsilon \proj{S_0},$
which is the convex sum of the maximally mixed state and the singlet state $S_0 = (\ket{01}-\ket{10})/\sqrt{2}$.  While $\rho_W$ is quantum correlated and has nonvanishing discord for all values $\epsilon > 0$, it is entangled and has nonzero negativity for $\epsilon > 1/3$ \cite{PhysRevA.86.012309}.  Here the RS is interesting as the initial state $\rho_-$ is quantum uncorrelated and therefore the predictions correspond to the amount of discord or negativity in $\rho_W$ itself.  
Fig. \ref{werner} plotting the predicted values $\Delta N$ and $\Delta D$ versus the purity factor $\epsilon$ shows an excellent agreement between actual and predicted values. RMSD also remained less than 0.05 indicating the high quality of predictions.

\subsection{Dependence on the dimensions of database and latent vector}
\begin{figure}
	\centering
	\includegraphics[trim=2.9cm 11.7cm 1.5cm 11.7cm,width=8.5cm,clip=]{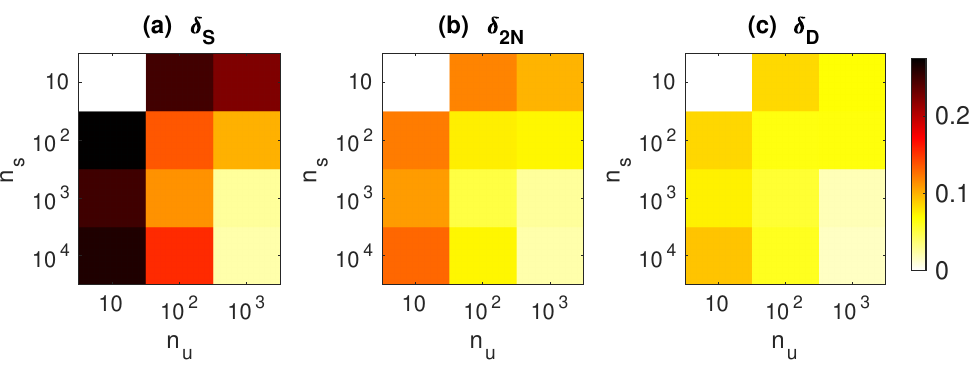}
	\caption{RMSD values of entropy (a), negativity (b), and discord (c), for various sizes of the database, but with fixed number $n_r = 100$ of unknown ratings.}
	\label{nsnu}
\end{figure}

Here we consider the two-qubit case.
We can fix the number of unknown ratings $n_r = 100$, and vary the dimension of the database.  We constructed a set of databases with number of states $n_s$ varying from 10 to 10,000, and the number of unitaries $n_u$ varying from 10 to 1000.  Fig. \ref{nsnu}(a-c) display RMSD values   of entropy, negativity, and discord respectively.  As expected, for a fixed number $n_r$ of unknown ratings, the RS efficiency in terms of RMSD  improves with the size of the database.  We observe an interesting diagonal symmetry indicating better efficiency with roughly equal number of states and evolutions.
It is also interesting to note that while the entropy rating is the most sensitive to the database size, the discord rating is least sensitive.  

\begin{figure}
	\centering
	\includegraphics[trim=3.7cm 11cm 2cm 11.7cm,width=8cm,clip=]{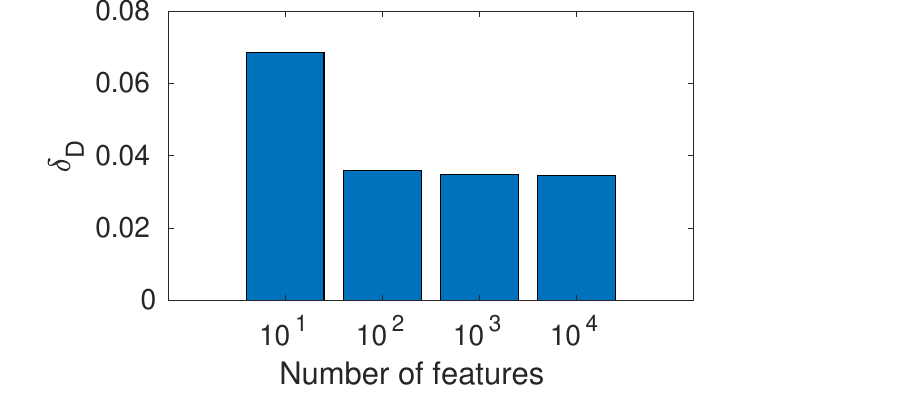}
	\caption{RMSD values of discord predictions versus number of features ($f$) for 50\% sparse 1000$\times$1000 database.}
	\label{varynf}
\end{figure}

%\subsection{Dependence on the latent-vector dimension}
Since the latent-vectors capture the mathematical structure of the database and thereby help predicting the unknown ratings, the dimension of the latent space, i.e., number of features $f$, becomes very important.  In the following example (Fig. \ref{varynf}) we consider a $1000\times 1000$  two-qubit database of discord ratings with 50\% of elements being removed and predicted (i.e., $n_r = 500,000$).  On running the RS for varying number of $f$, we find that if the latent vectors are too short, the prediction error becomes too large, since the RS fails to capture the underlying rules of the database.  Too large latent vectors also render the RS over-determined and inefficient.  One can find the appropriate number of features with a few trial and error runs.  In this particular case, $f\sim 10^2$ would suffice.

\subsection{Identifying noisy database}
\begin{figure}
	\centering
	\includegraphics[trim=2.5cm 8.6cm 2.5cm 9.0cm,width=7.5cm,clip=]{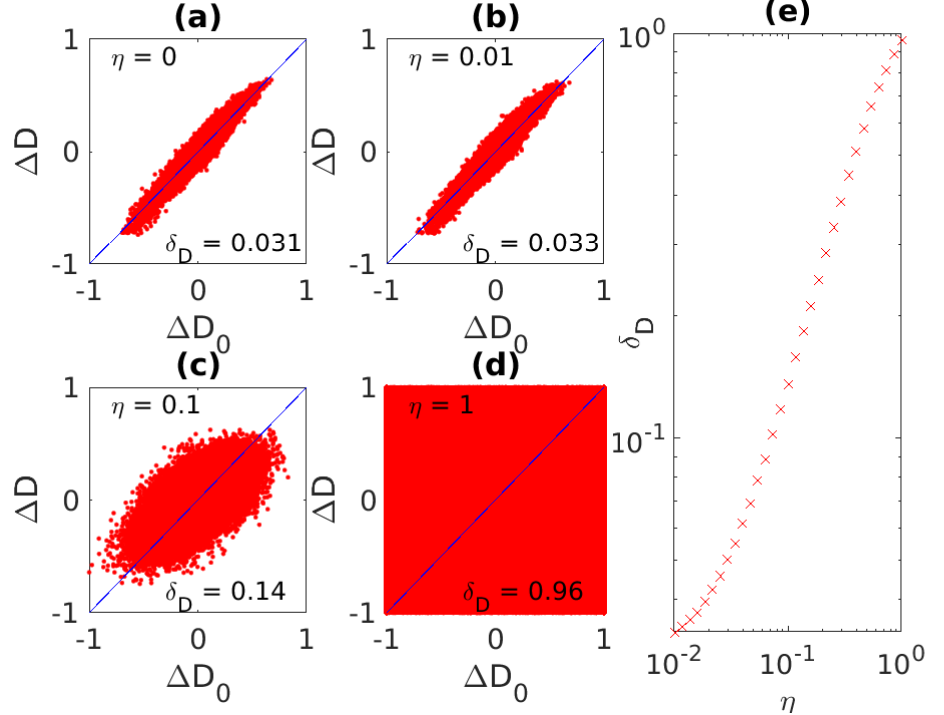}
	\caption{(a-d) Discord predictions with varying noise parameter $\eta$. (e) RMSD  $\delta_D$ versus $\eta$.  The number of unknowns $n_r = 500,000$, which is 50\% of the 1000$\times$1000 database.}
	\label{compnoise}
\end{figure}

What would be the dependence of prediction efficiency, if the input database itself is noisy, unreliable, or even fake? What if there are no underlying structures in the database so that no latent space exists? Can we use this dependency for asserting if a database, with or without a large number of unknown elements, is genuine or not?  In order to probe these questions, we choose discord rating in  a two-qubit register.  We first construct a $1000 \times 1000$ noise-free discord database ${\cal R}^{(0)}$, from which we setup a noisy database
${\cal R}^{(\eta)} = \eta S +(1-\eta) {\cal R}^{(0)},$ 
where $S$ is a random matrix of same dimensions and the noise parameter $\eta \in [0,1]$.  Thus, as $\eta$ discretely varies from 0 to 1, the corresponding database goes from a clean discord database to a random matrix.  In each database, corresponding to a fixed value of $\eta$, we randomly remove 500,000 entries constituting 50\%  of elements.  We now attempt to predict the missing elements starting from each of these databases.  The results shown in Fig. \ref{compnoise} indicates that for small noise parameter $\eta \le 0.01$, there is no significant effect on the RMSD value.  However, as the noise builds up, the RS finds it increasingly difficult to rate the entries, and RMSD increases by about five times for 10\% noise.  Finally, as expected, the RS completely fails for the completely random database corresponding to $\eta = 1$.  Thus, if a control database is provided, the RS can help assess the quality of a target database.

\subsection{Computational Time}
\begin{figure}
	\centering
	\includegraphics[trim=4.3cm 6.6cm 5.7cm 6.8cm,width=9cm,clip=]{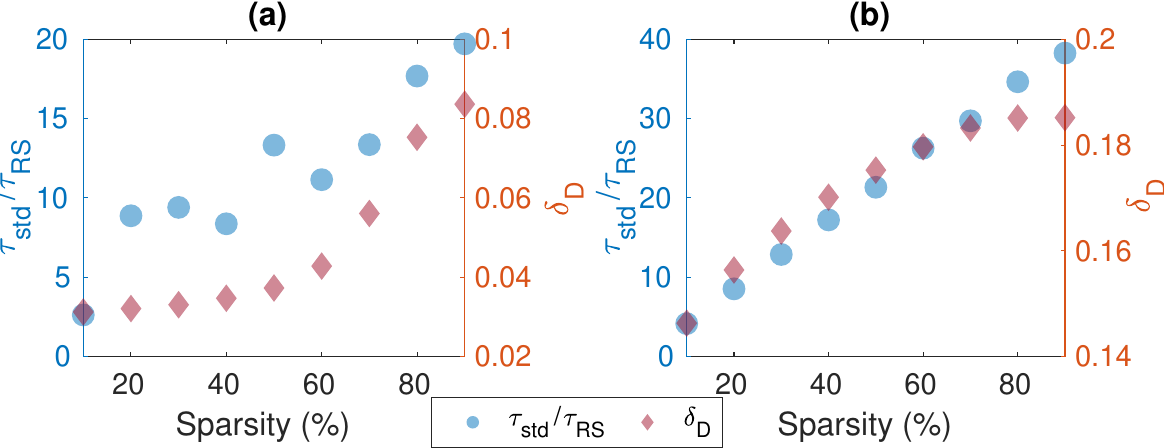}
	\caption{Computational advantage  $\tau_\mathrm{std}/\tau_\mathrm{RS}$ (left-vertical axes) and RMSD $\delta_D$ (right-vertical axes) vs database sparsity (in this case $n_r/10000$) for  two- (a) and three-qubit (b) registers.}
	\label{comptime}
\end{figure}

Since computing quantum discord is \textit{hard}, it is interesting to compare the computational time  $\tau_{\mathrm{RS}}$ for the RS prediction of
quantum discord with the calculation time $\tau_\mathrm{std}$ by the standard approach 
that involves subtracting maximum possible classical correlation from total correlation \cite{PhysRevLett.88.017901,PhysRevA.86.012309}. 
First we consider a 1000$\times$1000 two-qubit database.  For a two-qubit system, an optimal set of measurements exists that substantially reduces computational time  for estimating the discord \cite{PhysRevA.83.012327}.  Even in this case, RS brings about a substantial advantage by factors ranging from 3 to 20 for 10\% to 90\% sparsity of  the database, while ensuring RMSD values below 0.1 (Fig \ref{comptime}(a)).  
Unlike the two-qubit case, no optimal measurement-sets are known for larger registers. Therefore we also analyze the RS computational efficiency in a 1000$\times$1000 three-qubit database (Fig \ref{comptime}(b)).  In this case,
the computational advantage is up to 40, i.e., almost doubled over the two-qubit case, albeit at the cost of higher RMSD values.  Thus RS predictions, within certain precision limits, can be much faster than the standard calculations.

\subsection{Predicting state fidelity} 
\begin{figure}
	\centering
	\includegraphics[trim=0.7cm 5cm 3.2cm 5cm,width=8.8cm,clip=]{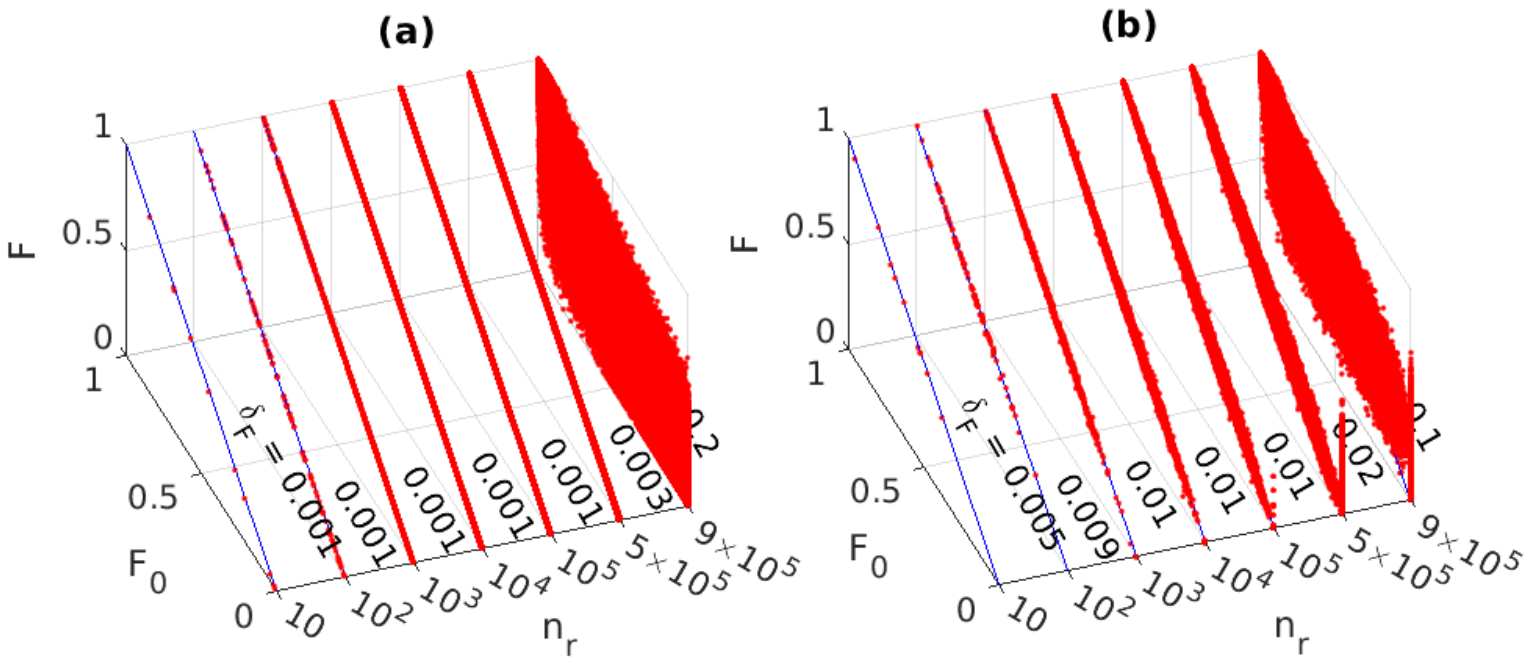}
	\caption{Predicted fidelity $F$ versus actual fidelity $F_0$ and the number $n_r$ of unknown elements for pure (a) and mixed (b) states.  Blue lines represent the ideal case $F = F_0$.  The RMSD value $\delta_F$ is shown in each case.}
	\label{fid}
\end{figure}

Like predicting quantum correlations, one can also predict the state-fidelity in the same manner.  For a set of pure input states $\{\ket{\psi_i}\}$ and unitary evolution $\{U_i\}$, the database elements 
${\cal R}_{i,j} = F_{i,j}(\psi_i,U_j)  = \vert \langle \psi_i \vert U_j \vert \psi_i \rangle \vert^2$
describe the fidelity of output state $U_j \ket{\psi_i}$ with the input state $\ket{\psi_i}$.  Similarly, for a set of mixed input states $\{\rho_i\}$ we may use the Uhlmann trace distance
${\cal R}_{i,j} = F_{i,j}(\rho_i,U_j) = \left\vert \mbox{Tr}\sqrt{\rho_i^{1/2} U_j \rho_i U_j^\dagger \rho_i^{1/2}} \right\vert^2$.
Of course, here instead of the input state, one could have also chosen any other target state.
The results of the fidelity predictions using a database of 1000 random states and 1000 unitary evolutions with $n_r$ unknown elements are shown in Fig. \ref{fid}.  Interestingly, the rating is extremely successful for pure states (Fig. \ref{fid}(a)), for which up to 50\% of unknown elements ($n_r = 5\times 10^5$) can be predicted with a low RMSD $\delta_F \le 0.003$.  For mixed states the RMSD values are relatively higher, but still $\delta_F \le 0.02$ for predicting up to 50\% of unknown elements.  However, in both cases the predictions fail for 90\% of unknown elements.

\subsection{Identifying local and nonlocal operators:}
\begin{figure}
	\centering
	\includegraphics[trim=3cm 7cm 3cm 7cm,width=8.5cm,clip=]{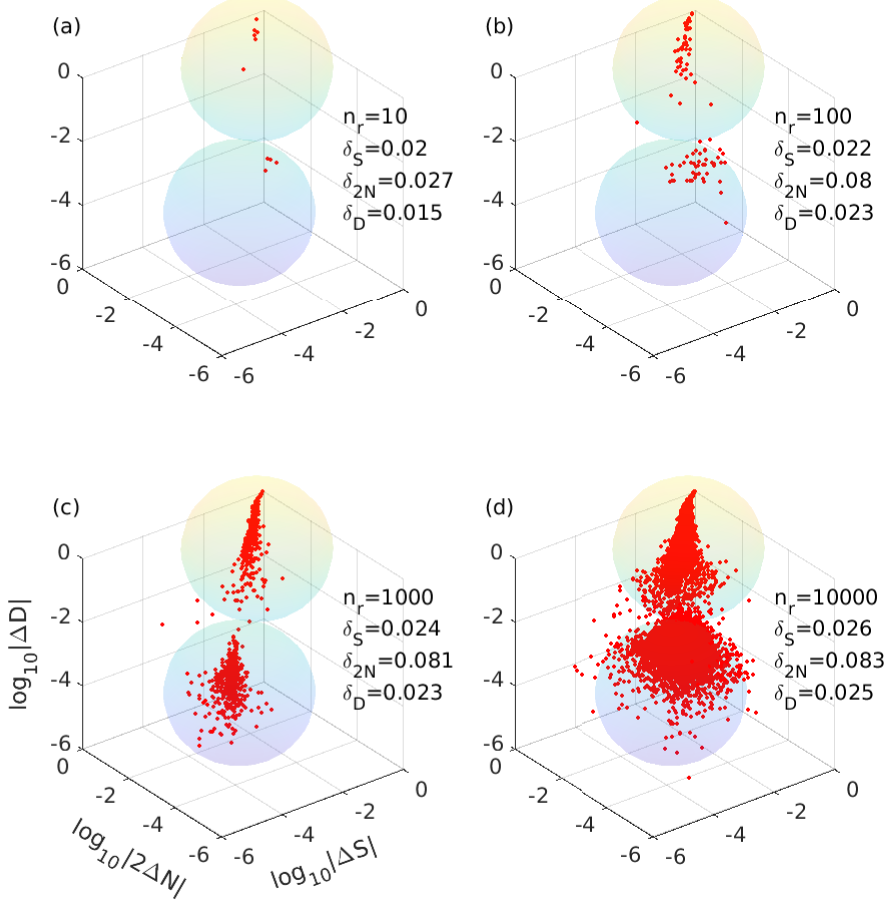}
	\caption{Binning unitaries into local or nonlocal by predicting the changes in quantum correlations.  Subplots correspond to different numbers $n_r$ of predicted ratings as indicated in each case. }
	\label{locnonloc}
\end{figure}

Suppose the objective is not to obtain quantitative values for the changes in quantum correlations introduced by unitary operators, but rather to identify local and nonlocal unitary operators.  For this purpose, we generate a database with 1000 randomly generated unitaries, 500 of which are by construction local operators of the form $U_A \otimes U_B$ that do not change quantum correlations.  The remaining $4\times 4$ dimensional random unitaries $U_{AB}$ are mostly nonlocal and introduce changes in quantum correlations. These propagator act on 1000 randomly generated initial states, so that the overall size of the database is $10^6$.
Fig. \ref{locnonloc} shows the three-dimensional simultaneous plots of predicted  entropy, negativity, and discord values for various numbers $n_r$ of unknown ratings. It is clear that most of the points fall into one of the two bins - negligible correlation changes resulting from local unitaries, or significant correlation changes brought about by nonlocal unitaries.

\subsection{Scaling to larger registers}
\label{multi}
\begin{figure}[t]
	\centering
	%	\hspace*{-0.4cm}
	\includegraphics[trim=1.4cm 2.6cm 1.6cm 2.5cm,width=8cm,clip=]{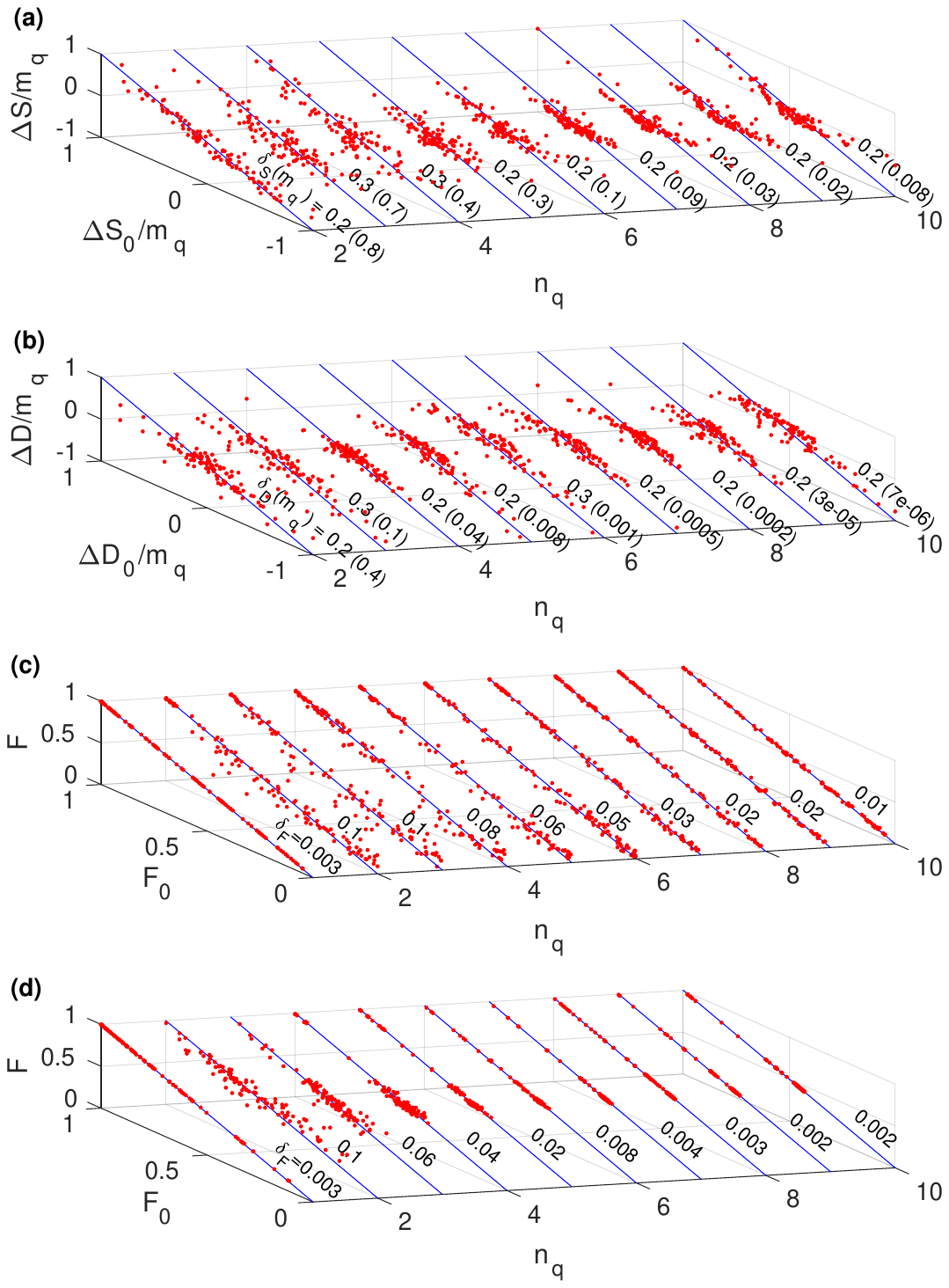}
	\caption{RS predictions of (a) entropy, (b) quantum discord, (c) fidelity of pure states, and (d) fidelity of mixed states, for registers with $n_q$ qubits.  The RMSD values $\delta$ are shown in each case and the maximum range $m_q$ (in parenthesis) are shown in (a) and (b).}
	\label{corrfidnq}
\end{figure}

%\subsubsection{Correlation rating of multi-qubit unitaries}
We now study the efficiency of the RS in larger quantum registers, with sizes up to 10 qubits.  For each register, we constructed a database with 100 random unitary operators and 100 random quantum states.  The number of unknown ratings was fixed at $n_r = 100$.  In larger registers, there are multiple ways of partitioning the system for estimating quantum correlations.  For the sake of simplicity, we first partition the whole system of $n_q$ qubits into two parts: a two-qubit part $P$ and $n_q-2$ qubit part $Q$.  We estimated quantum correlations in the two-qubit part $\rho^P = \mbox{Tr}_Q[\rho^{PQ}]$, after tracing out $Q$.
Owing to the small subspace selected, the absolute change of correlation decreases rapidly with the number of qubits.  
%This effect was more severe in the case of negativity, which mostly vanished in registers with more than three qubits, and hence we now omit it from the discussions.  
For comparison purposes, we use the rescaled correlation change $\Delta_C/m_q$, with $m_q = \max\{|\Delta C_{0_{i,j}}|, |\Delta C_{i,j} |\}$ where maximum is taken over all the $n_r$ rated elements corresponding to the $n_q$-qubit register.
Fig. \ref{corrfidnq} displays the predicted values of changes in entropy (Fig. \ref{corrfidnq}(a))  as well as discord (Fig. \ref{corrfidnq}(b)) versus the actual values for various sizes of quantum registers with up to 10 qubits.  Here the RMSD values are calculated with respect to the rescaled correlation changes.  Despite the smaller size of database, the predictions were largely in agreement with the actual values, and the RMSD values mostly remained below 0.3, thus confirming the feasibility of the RS predictions for various system sizes.

%\subsubsection{Fidelity rating of multi-qubit unitaries}
We have also studied the fidelity rating in multi-qubit registers.  Fig. \ref{corrfidnq} also displays the results of the fidelity rating of pure (Fig. \ref{corrfidnq}(c)) and mixed (Fig. \ref{corrfidnq}(d)) states in registers with up to 10 qubits.  In each case, we used a database of 100 states and 100 unitaries, while the number of unknown entries was fixed to 100.  It is clear that the fidelity rating is largely successful in all these cases.

\begin{figure}
	\centering
	\includegraphics[trim=0.4cm 4cm 1.0cm 3.5cm,width=8.5cm,clip=]{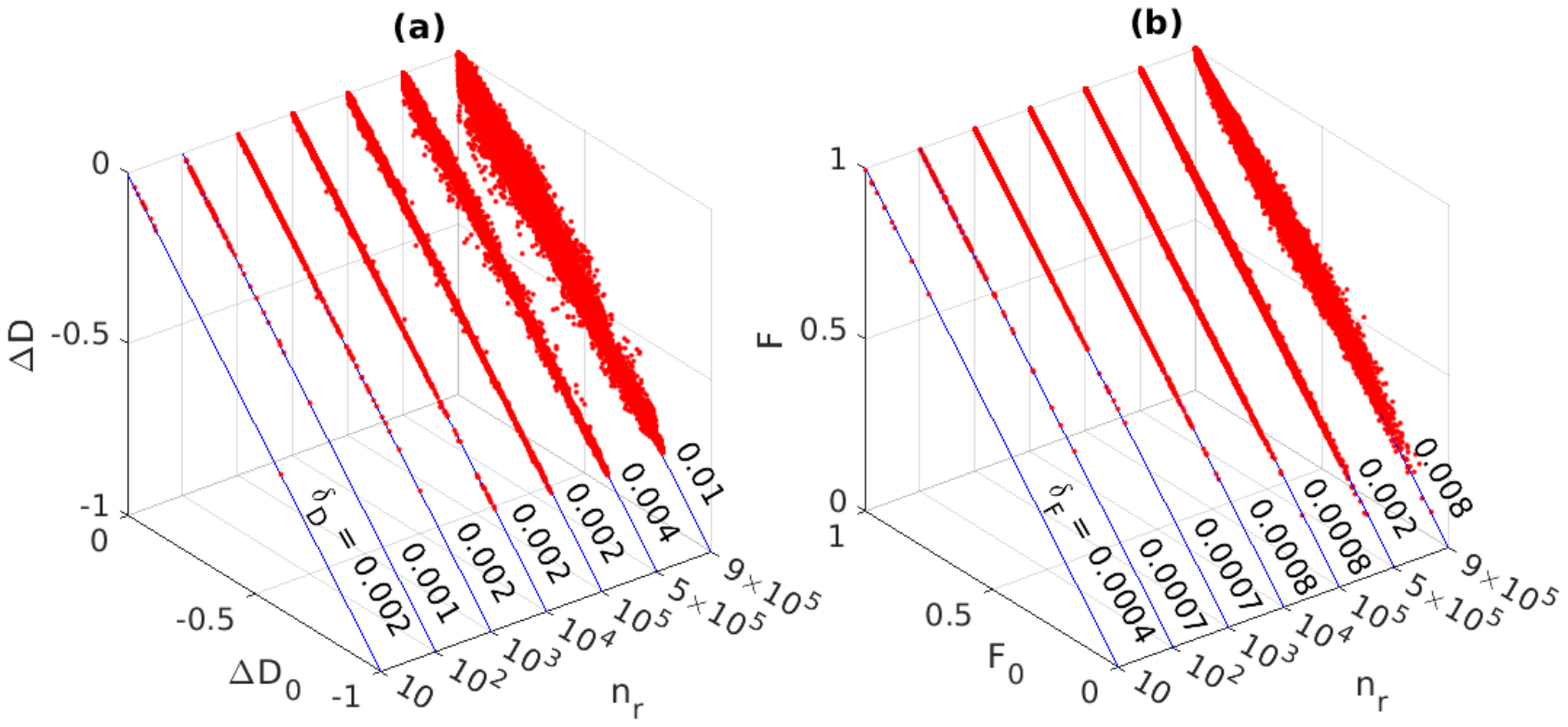}
	\caption{The RS predictions of nonunitary evolutions.  Predicted values (red points) for changes in discord (a) and fidelity (b) versus actual values and the number $n_r$ of unknown elements.  In each case, the size of the database is $1000\times1000$.  Blue lines represent the ideal curves.  The RMSD values are shown in each case.}
	\label{nonunitary}
\end{figure}

\section{Rating nonunitary evolutions}
\label{nonunitarysec}
Here we consider a noisy two-qubit system with independent single-qubit decoherence channels each of which having an operator-sum representation in terms of Kraus operators $\{M_k\}$ \cite{nielsen2002quantum}.
An input state $\rho_i^{AB}$  transforms according to 
\begin{align}
		{\cal E}_j(\rho_i^{AB}) &= \frac{1}{2}\sum_k (M_k^{j_A} \otimes \mathbbm{1}) \rho_i^{AB} (M_k^{j_A^\dagger} \otimes \mathbbm{1}) \nonumber \\
		&+ \frac{1}{2}\sum_l (\mathbbm{1} \otimes M_l^{j_B}) \rho_i^{AB} (\mathbbm{1} \otimes M_l^{j_B^\dagger}), 
	\end{align}
%wherein channels $j_A$ and $j_B$ on qubits $A$ and $B$ respectively.  We consider the scenario 
wherein the single-qubit channels $j_A$ and $j_B$ belong to one of the following channels: bit-flip, phase-flip, bit \& phase-flip, depolarization, and amplitude damping as shown below.
\begin{itemize}
	\item[(i)] Bit-flip:
	$M_1^\mathbbm{X} = \sqrt{x} ~\mathbbm{1},  ~
	M_2^\mathbbm{X} = \sqrt{1-x}~\sigma_x,$
	\item[(ii)] Phase-flip:
	$M_1^\mathbbm{Z} = \sqrt{z} ~\mathbbm{1},  ~
	M_2^\mathbbm{Z} = \sqrt{1-z}~\sigma_z,$
	\item[(iii)] Bit \& phase-flip:
	$M_1^\mathbbm{Y} = \sqrt{y} ~\mathbbm{1},  ~
	M_2^\mathbbm{Y} = \sqrt{1-y}~\sigma_y,$
	\item[(iv)] Depolarization: 
	$M_1^\mathbbm{D} = \sqrt{1-\frac{3d}{4}}~\mathbbm{1},~
	M_2^\mathbbm{D} = \frac{\sqrt{d}}{2}~\sigma_x, \\
	M_3^\mathbbm{D} = \frac{\sqrt{d}}{2}~\sigma_y,~
	M_4^\mathbbm{D} = \frac{\sqrt{d}}{2}~\sigma_z,$ and
	\item[(v)] Amplitude damping: \\
	$M^\mathbbm{A}_1 = \left[ 
	\begin{array}{cc}
		1 & 0 \\
		0 & \sqrt{1-a}
	\end{array}
	\right],
	M^\mathbbm{A}_2 = \left[
	\begin{array}{cc}
		0 & \sqrt{a}\\
		0 & 0
	\end{array}
	\right]$. 
\end{itemize}
Here $x,y,z,a,d$ are randomly chosen channel probabilities and $\sigma_x$, $\sigma_y$, and $\sigma_z$ are Pauli operators.

%\begin{figure}
%	\centering
%	\includegraphics[trim=0.4cm 4cm 1.0cm 3.5cm,width=9cm,clip=]{nonunitary2q3drow.pdf}
%	\caption{The RS predictions of nonunitary evolutions.  Predicted values (red points) for changes in discord (a) and fidelity (b) versus actual values and the number $n_r$ of unknown elements.  In each case, the size of the database is $1000\times1000$.  Blue lines represent the ideal curves.  The RMSD values are shown in each case.}
%	\label{nonunitary}
%\end{figure} 

We form a random database of 1000 mixed two-qubit states and 1000 nonunitary evolutions, with a particular pair of channel probabilities, acting independently on the individual qubits.  Fig. \ref{nonunitary} displays the predicted values of change in discord
$\Delta D_{{i,j}} = D({\cal E}_j(\rho_i))-D(\rho_i)$
(Fig. \ref{nonunitary}(a)) and of Fidelity
$ F_{i,j}(\rho_i,{\cal E}_j) = \left\vert \sqrt{\sqrt{\rho_i} {\cal E}_j(\rho_i)\sqrt{\rho_i}} \right\vert^2$
(Fig. \ref{nonunitary}(b)), plotted versus the actual values for various numbers $n_r$ of unknown elements.  Since the decoherence channels only reduce quantum correlations, $\Delta D$ is essentially negative.  It is evident that the prediction is highly  successful, with the RMSD values being 0.01 or lesser even at 90\% of unknown elements.

\section{Construction of quantum phase space}
\label{sec:application}

	As the demonstration of a specific application of RS predicted quantum correlation database, we now describe an efficient construction of discord phase space  for studying quantum chaos.  Implications of quantum chaos for quantum information processing is well studied \cite{PhysRevE.62.6366, Hauke_2012}. Classical chaos is often studied by observing the trajectory of a dynamical system in a certain phase space \cite{haake1987classical}.  It has been shown that quantum discord phase space can be used to study quantum chaos of a dynamical quantum system \cite{PhysRevE.91.032906}.  However, the discord calculation is not based on a compact analytical expression, instead requires expensive computation overhead.  In the following, we show that RS can efficiently generate discord phase space for quantum dynamical systems, such as a quantum kicked top (QKT) with the Hamiltonian
	\begin{equation}
		H = \frac{\kappa}{2j\tau} J_z^2 + \frac{\pi}{2} J_x \sum_{n = 0}^{\infty} \delta(t-n\tau).
	\end{equation} 
	Here $\boldsymbol{J} = [J_x, J_y, J_z]$ is the angular momentum vector of an effective spin-$j$ system. The first term represents the nonlinear evolution for a duration $\tau$ with chaoticity  parameter  $\kappa$. 
	%For a given value of j, system can be decomposed in N = 2j qubits where hilbert space dimension is 2j+1. 
	The second term represents periodic linear kicks.
	% where f is rectangular pulse with duration $\Delta$ ($\Delta<<\tau$) and strength p.  Here we set $p\Delta = \frac{\pi}{2}$.\\
	The time evolution is governed by the Floquet propagator
	\begin{equation}
		U_{QKT} = e^{-i (\pi/2) J_x} e^{-i \kappa J_z^2/(2j)} .
	\end{equation}

	We model the QKT system using 2 qubits.  Starting from an initial state $\ket{00}$,
	%	given by 
	%\begin{equation}
	%	\ket{j,j} = \ket{00}
	%\end{equation}
	we prepare the spin coherent state $\ket{\theta,\phi}\otimes\ket{\theta,\phi}$, where
	%in the hilbert space can be given by using rotation matrix $R(\theta, \phi)$ such as 
	\begin{equation}
		\ket{\theta,\phi} = \cos(\theta/2)\ket{0}+e^{i\phi}\sin(\theta/2)\ket{1}
		%	\exp{-i\theta J_x}sin\phi - J_y cos \phi)] \ket{j,j}
	\end{equation}
	by first rotating each qubit with an angle $\theta$ about the y-direction followed by an angle $\phi$ about the z-direction in the Bloch sphere.
	%using the rotation $R(\theta, \phi) = ; 0 \le \theta \le \pi, 0 \le \phi \le 2\pi $

\begin{figure}[t]
	\centering
	\includegraphics[trim=1.5cm 2cm 0.8cm 2.3cm,width=9cm,clip=]{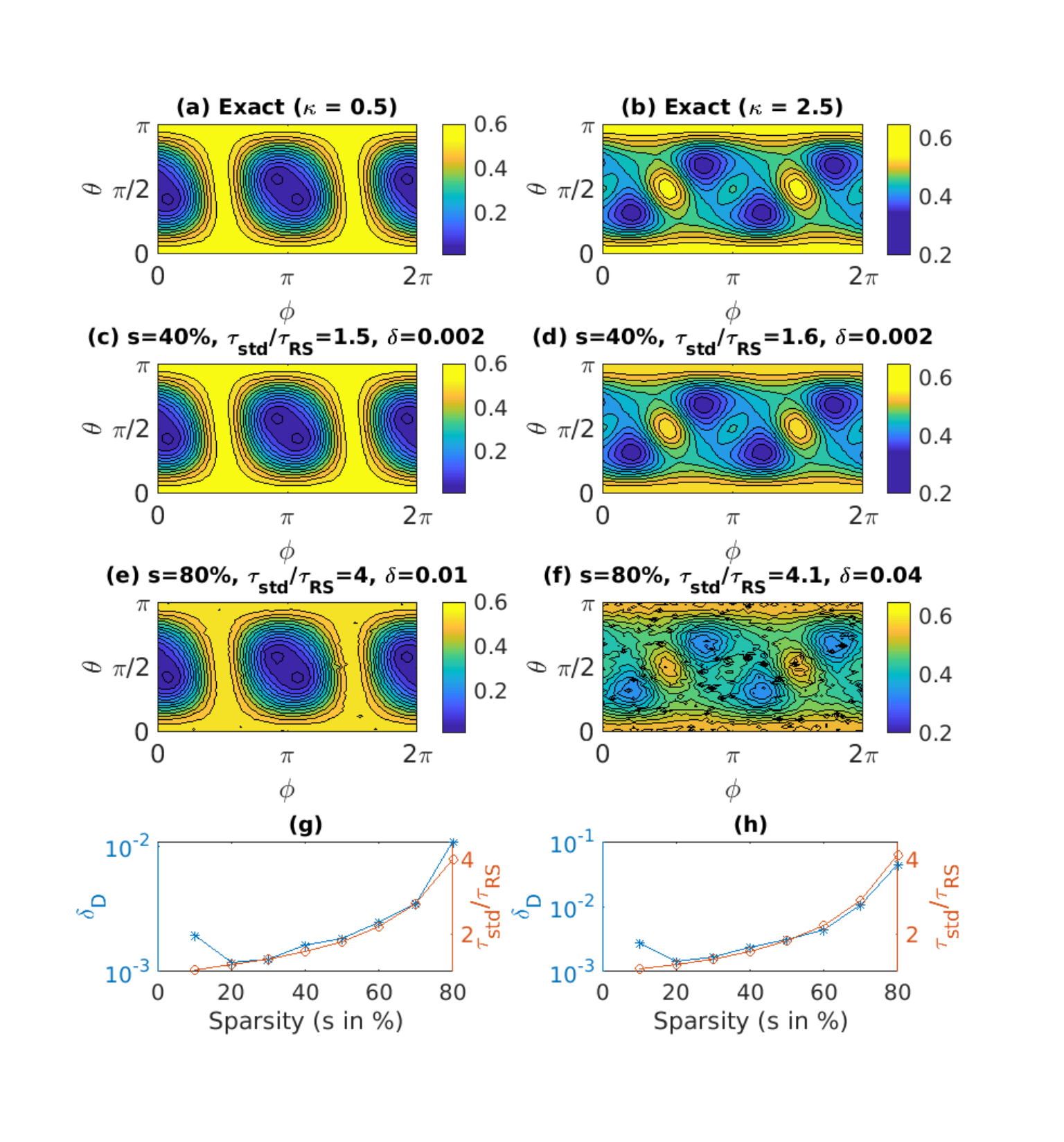}
	\caption{Discord phase space plots (a-f) for chaoticity parameter $\kappa = 0.5$ (first column) and for $\kappa = 2.5$ (second column).
			The exact phase space plots in (a) and (b) are constructed with standard method, while the other phase space plots are predicted by RS method with sparsity value of either 40\% (c,d) or 80\% (e,f).  The RMSD values ($\delta_D$) quantifying the mismatch between the exact and the RS predicted phase space plots, as well as the corresponding time advantage factors ($\tau_\mathrm{std}/\tau_\mathrm{RS}$) are also plotted against varying sparsity values (g,h).}
	\label{application}
\end{figure}

Fig. \ref{application} shows discord phase space plots of the two-qubit QKT for two chaoticity values viz., $\kappa = 0.5$ (first column) and $\kappa = 2.5$ (second column) obtained by averaging over 100 kicks with a $51\times 51$ grid of $\theta$ and $\phi$ coordinates.  We compare the exact discord phase space diagrams (Fig. \ref{application} (a,b)) with those predicted by RS (c-f) using a total feature number $n_f = 100$.	As expected, at low chaoticity value of $\kappa = 0.5$, the QKT mostly undergoes a regular dynamics and the phase space exhibits large regions of closed trajectories. However, at a higher chaoticity value of $\kappa = 2.5$, it undergoes a rather complex dynamics whose phase space exhibits intricate patterns with smaller regular islands surrounded by chaotic ocean.  It is evident that for $\kappa = 0.5$, RS predicted phase space plots are almost identical with the exact plot even for sparsity value of 80\%.  It is interesting to see that at higher chaoticity value of $\kappa=2.5$, RS prediction matches well for the sparsity value of $40\%$, while worsens for $80\%$.  This further reinforces the idea that the chaotic regimes of dynamical systems are hard to predict in general. The sensitivity of QKT to experimental imperfections in such chaotic regimes has been reported earlier by Krithika et al. \cite{PhysRevE.99.032219}.  Hence, it is natural that RS finds it challenging to recognize the underlying patterns in the database, based on which it can predict the missing entries. In spite of this difficulty, the overall predicted phase space pattern with $80\%$ sparsity is qualitatively similar to the exact one. The RMSD values ($\delta_D$) between the exact and the predicted plots, along with the computational time-advantage ($\tau_\mathrm{std}/\tau_\mathrm{RS}$) are plotted versus sparsity values in Fig. \ref{application} (g,h).  While there is a trade off between the prediction error and the time-advantage, it is clear that the RS prediction is quite robust and efficient, and is about four times faster for 80\% sparsity. The advantage is likely to be even higher for larger dimensional quantum systems.

\section{Conclusions}
\label{summary}
We have adapted a class of recommender systems, called the matrix factorization algorithm, to characterize quantum evolutions  in terms of quantum correlations as well as state-fidelity. First, using two-qubit databases, we have carried out a detailed analysis of the recommender system predicting three types of correlations, namely  entropy, negativity, and discord.  We found that the recommender system was able to efficiently predict well over 50\% of unknown elements in the database.
As a particular example, we showed that predictions of negativity and discord of Werner state prepared with different purities matched very well with the expected values.  
By introducing noise into the database in a systematic way, we observed the prediction efficiency deteriorating with noise.  We proposed that this fact, along with a reference database, could be used to distinguish a genuine or clean database form a noisy or a fake database.
Predicting quantum discord of a general state, for which a closed-form expression does not exist, is most interesting.  The analysis of computational time shows that the machine prediction, within certain precision limits, can be far more efficient. In million-element databases of two- and three-qubit registers, we observed over an order of magnitude improvement in computational time.  To show that the characterization of quantum evolutions via recommender system is scalable with system size, we have demonstrated predicting quantum correlations as well as fidelities of unitary evolutions on larger multiqubit registers with up to ten qubits.
We have also demonstrated the capability of  the recommender system in characterizing nonunitary evolutions by subjecting it to predict correlation and fidelity ratings of states undergoing certain decoherence channels.
Finally, we demonstrate a robust RS prediction of  discord phase space diagrams useful to study quantum chaotic systems. 

When does the RS fail? We have earlier discussed the noisy database having no underlying pattern and thus fails to converge.  RS can fail in many other scenarios.  For example, the RS can not predict elements of an empty row or column of a database.  We have demonstrated that for a fixed number of unknowns, the prediction error gets worse by shrinking the database along any dimension.  Rating a state (or gate) that is far away from all other members of the database may be difficult.  We also found that too short latent vectors also fail to capture the mathematical structure beneath the database. For an intricate database, like that of long-range quantum correlations, one  needs larger latent vectors and more powerful global optimization methods.  

Although here we have only discussed databases of random states and random evolutions, it is straightforward to adapt it to other interesting scenarios such as quantum process tomography \cite{PhysRevLett.78.390} and in predicting quantum correlations in a time-evolving many-body system initialized with various possible states.  It would also be interesting to predict the behavior of quantum systems under evolutions dictated by more sophisticated master equations.  Further improvisations in the precision and speed of machine predictions will go a long way as well. The remarkable efficiency of the recommender system in characterizing quantum evolutions hints at an underlying mathematical structure that may deserve further exploration.
We anticipate that such machine learning techniques not only aid the evolution of quantum technologies but also provide deeper insights into mathematical structures in quantum theory itself.

\section*{Acknowledgments}
Authors thank Prof. Santhanam, Dr. G. J. Sreejith, Krithika, and Soham for useful discussions. P. B. gratefully acknowledges support from the Prime Minister's Research Fellowship and the discussion meeting Statistical Physics of Machine Learning (ICTS/SPMML2020/01) at International Centre for Theoretical Sciences Bangalore for all useful discussions. A. S. gratefully acknowledges support from the Summer Program-2019 at IISER Pune. T.S.M. acknowledges funding from DST/ICPS/QuST/2019/Q67.

\bibliographystyle{unsrtnat}
\bibliography{ref}

\end{document}